\documentclass[referee,a4paper,12pt,traditabstract]{swsc} 

\usepackage{graphicx}
\usepackage{txfonts}
\usepackage{subfigure}
\usepackage{epstopdf}
\usepackage{lineno}
\usepackage[authoryear,round]{natbib}
\usepackage[backref]{hyperref}
\usepackage{url}

\hypersetup{colorlinks=true,citecolor=cyan,urlcolor=cyan,linkcolor=blue}

\begin{document}


   \title{Investigating the Kinematics of Coronal Mass Ejections with the Automated CORIMP Catalog}
   
   \titlerunning{Investigating the Kinematics of CMEs with the Automated CORIMP Catalog}

   \authorrunning{J. P. Byrne}

   \author{Jason P. Byrne}

   \institute{RAL Space, Rutherford Appleton Laboratory, Harwell Oxford, OX11 0QX, UK. \\
              \email{\href{mailto:jason.byrne@stfc.ac.uk}{jason.byrne@stfc.ac.uk}}
                      }


 
  \abstract
  {
  Studying coronal mass ejections (CMEs) in coronagraph data can be challenging due to their diffuse structure and transient nature, compounded by the variations in their dynamics, morphology, and frequency of occurrence. The large amounts of data available from missions like the Solar and Heliospheric Observatory (SOHO) make manual cataloging of CMEs tedious and prone to human error, and so a robust method of detection and analysis is required and often preferred. A new coronal image processing catalog called CORIMP has been developed in an effort to achieve this, through the implementation of a dynamic background separation technique and multiscale edge detection. These algorithms together isolate and characterise CME structure in the field-of-view of the Large Angle Spectrometric Coronagraph (LASCO) onboard SOHO. CORIMP also applies a Savitzky-Golay filter, along with quadratic and linear fits, to the height-time measurements for better revealing the true CME speed and acceleration profiles across the plane-of-sky. Here we present a sample of new results from the CORIMP CME catalog, and directly compare them with the other automated catalogs of Computer Aided CME Tracking (CACTus) and Solar Eruptive Events Detection System (SEEDS), as well as the manual CME catalog at the Coordinated Data Analysis Workshop (CDAW) Data Center and a previously published study of the sample events. We further investigate a form of unsupervised machine learning by using a $k$-means clustering algorithm to distinguish detections of multiple CMEs that occur close together in space and time. While challenges still exist, this investigation and comparison of results demonstrates the reliability and robustness of the CORIMP catalog, proving its effectiveness at detecting and tracking CMEs throughout the LASCO dataset.
  }

   \keywords{Sun -- Coronal Mass Ejection (CME) -- Space weather -- Solar image processing -- Machine learning}

   \maketitle

\section{Introduction}

Coronal mass ejections (CMEs) represent the largest, most dynamic phenomena that originate from the Sun \citep{2011LRSP....8....1C,2012LRSP....9....3W}. Propagating at speeds of up to thousands of kilometres per second, with energies on the order of 10$^{32}$\,ergs, they can drive adverse space weather throughout the solar system \citep{2005A&A...440..373H,2007LRSP....4....1P}. Given their potentially hazardous impact on Earth's geomagnetic environment, the physics governing their eruption and propagation needs to be understood so that their effects may be predicted in the guise of space-weather forecasting. To this end, observations of CMEs must be rigorously inspected in order to determine their dynamics, and this is most commonly undertaken with the use of coronagraph instruments \citep[e.g.,][]{1975Koomen,1980ApJ...237L..99S,1980SoPh...65...91M,1985JGR....90..275I,1993JGR....9813177H, 1995SoPh..162..357B, 2008SSRv..136...67H}. 

CMEs tend to be faint, transient phenomena, observed in white-light images that are prone to noise and user-dependent biases in their interpretation. During solar minimum, CMEs can occur every few days, but at solar maximum there can be several per day \citep{2000JGR...10518169S, 2004JGRA..10907105Y}. They exhibit a wide variety of morphologies, moving in unpredictable directions and speeds in the solar wind \citep{angeo-27-4491-2009, 2010NatCo...1E..74B, 2014NatCo...5E3481L}. They can drive shocks in the solar atmosphere and interplanetary space \citep{2005A&A...440..373H, 2013NatPh...9..811C}, and exhibit various levels of geo-effectiveness \citep{2001JASTP..63..389P, 2005AnGeo..23.1033S, 2009GeoRL..3608102D, 2012SoPh..tmp...47L}. A wealth of image processing techniques have been explored to study CMEs in remote-sensing image data provided by such instruments as the Large Angle Spectrometric Coronagraph \citep[LASCO;][]{1995SoPh..162..357B} onboard the Solar and Heliospheric Observatory \citep[SOHO;][]{1995SoPh..162....1D}. These techniques generally rely on some form of image differencing to highlight moving features in the observed intensities, but this introduces spatiotemporal crosstalk and scaling issues that affect the accuracy of CME characterisations. For example, the distance a CME moves between frames of varying cadence, along with its inherent morphological and brightness changes during that time, directly affects the calculations of running-difference images - to the point of changing how a user or algorithm characterises the CME structure and consequently its dynamics. More advanced image processing methods have thus been explored, such as optical flow techniques \citep{2006ApJ...652.1747C}, supervised segmentation techniques \citep{Goussies:2010:DTC:1749630.1750150}, wavelets \citep{2003A&A...398.1185S} and curvelets \citep{2011AdSpR..47.2118G}. The large volume of data available has made it necessary to automate such techniques for detecting and tracking CMEs across images, with a view to cataloging their kinematics and morphologies. This allows for more robust CME detections by avoiding the troublesome effects of standard image differencing techniques. It is therefore possible to maintain a non-biased characterisation of the CME structure in every event, since automated techniques are self-consistent.

To date, a point-and-click catalog of CMEs in LASCO data has been undertaken at the Coordinated Data Analysis Workshop (CDAW) Data Center \citep{2009EM&P..104..295G}, which operates by tracking CMEs in running difference images to produce information on the dynamics of each event. The CDAW Catalog is produced by a manual procedure and is therefore subject to user bias in interpreting the data. Automated catalogs have since been developed to overcome this bias and tedium. The Computer Aided CME Tracking routine \cite[CACTus;][]{2004A&A...425.1097R} is the first such automated catalog. It works by using a Hough transform \citep{Hough1962} to detect intensity ridges corresponding to CME tracks in time-height stacks (J-maps) of polar-unwrapped running-difference LASCO images. The Solar Eruptive Events Detection System \cite[SEEDS;][]{2008SoPh..248..485O} is another automated catalog that similarly uses polar-unwrapped running-difference LASCO images but with a form of threshold segmentation to approximate the shape of the CME leading edge in every image. A new automated catalog called CORIMP has recently been developed from a unique set of coronal image processing techniques, that overcomes many of the limitations of current catalogs in operation \citep{2012ApJ...752..144M,2012ApJ...752..145B}. An online database has been produced for the SOHO/LASCO data and event detections therein. It provides information on CME onset time, position angle, angular width, speed, acceleration, and mass, along with kinematic plots and observation movies. Such a wealth of information is crucial for understanding the dynamics of CMEs. Furthermore, a realtime version of the algorithm has been implemented to provide CME detection alerts to the interested space weather community\footnote{Automated email alerts may be requested from the author, and are also published on social media at \href{https://www.twitter.com/CMEcatalog}{twitter.com/CMEcatalog} and \href{https://www.facebook.com/CMEcatalog}{facebook.com/CMEcatalog}}.

In Section\,\ref{sect_catalogs} the CME catalogs are discussed in greater detail to highlight their methodologies and drawbacks. A sample of CMEs is then investigated in Section\,\ref{sect_events} in order to compare the outputs of each catalog, paying particular interest to the robustness and reliability of the new CORIMP catalog. In Section\,\ref{sect_clusters} a first effort is made to use a form of unsupervised machine learning to isolate spatially and temporally overlapping CME detections. The conclusions of this investigation are presented in Section\,\ref{sect_conclusions}.

\section{Cataloging CMEs}
\label{sect_catalogs}

\begin{figure}[!h]
\centerline{\includegraphics[width=\linewidth]{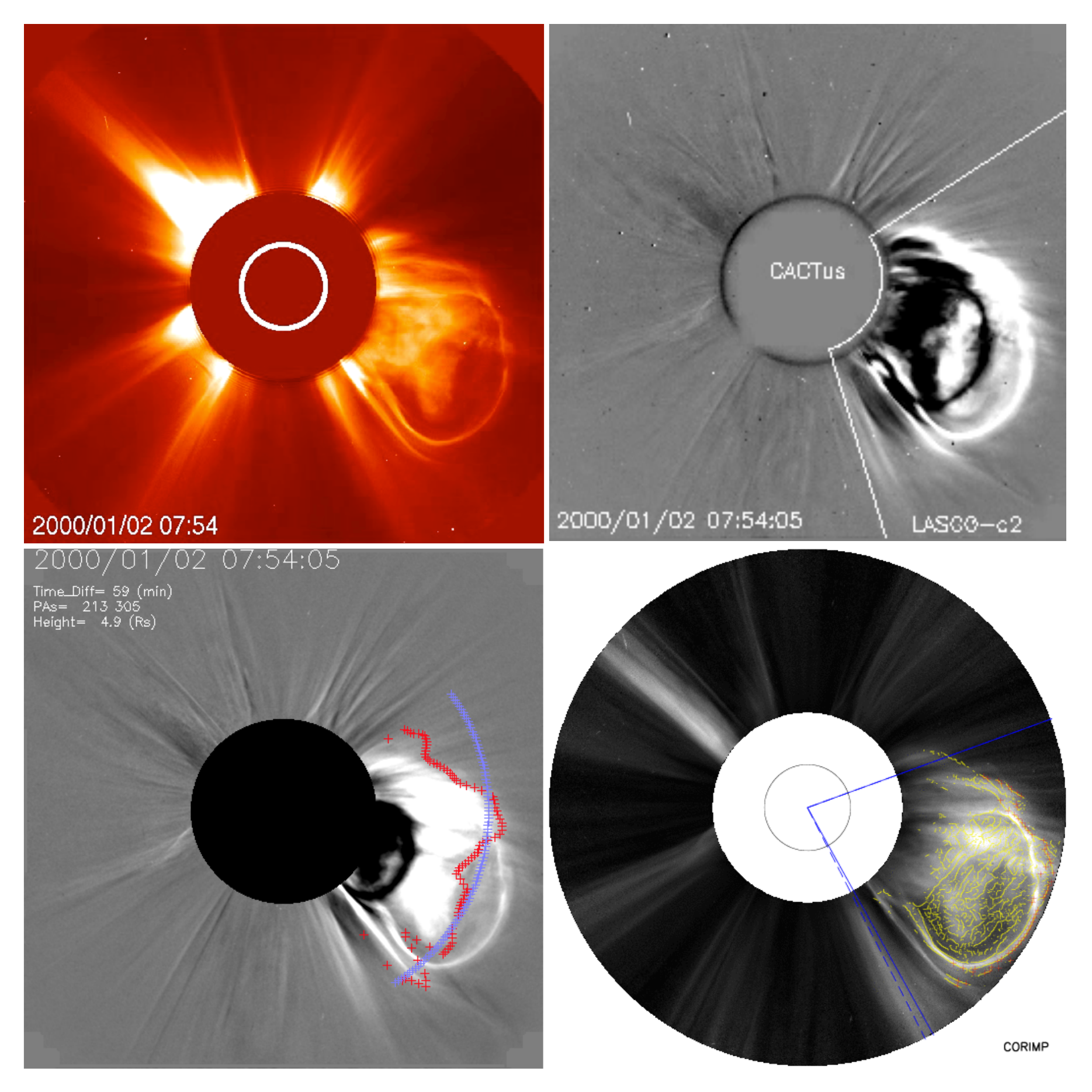}}
\caption{LASCO/C2 observations of a CME on 2000\,Jan.\,02 at 07:54\,UT. \emph{Top left:} A level 2 background-model subtracted image of the event. \emph{Top right:} Running difference image reproduced from the CACTus catalog with the angular span of the CME detection indicated by the white lines. (The manual CDAW catalog uses similar such running-difference images.) \emph{Bottom left:} Running difference image reproduced from the SEEDS catalog with the CME front detection highlighted in red (and the extended `half-max lead' in purple). \emph{Bottom right:} Normalised radial-gradient filtered (NRGF) image taken from the CORIMP catalog with the angular span of the CME detection indicated in blue, the pixel-chained CME structure in yellow, and the CME front in red.}
\label{20000102_four}
\end{figure}

\begin{figure}[!h]
\centerline{\includegraphics[scale=0.68, trim=20 0 0 0, clip=true]{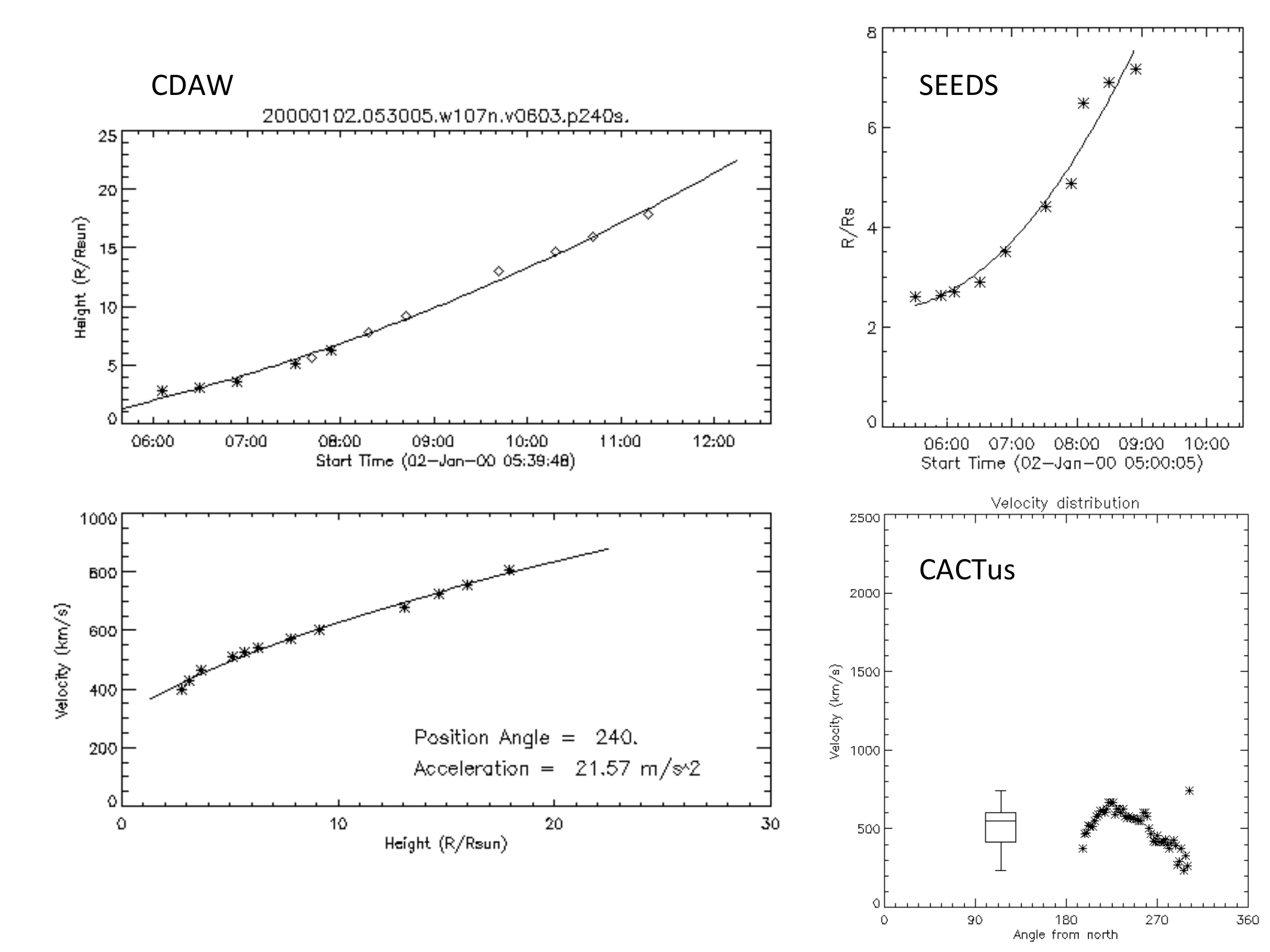}}
\caption{The kinematic outputs for the 2000\,Jan.\,02 CME reproduced from the CDAW, SEEDS and CACTus catalogs. \emph{Top left:} The CDAW catalog height-time measurements of the CME chosen manually along the running-difference bright front (at position angle 240) with a second-order fit. \emph{Bottom left:} The corresponding CDAW speed profile plotted against height, showing an acceleration of $21.57\,m\,s^{-2}$. \emph{Top right:} The automated SEEDS height-time measurements and second-order fit resulting in an acceleration of 18.6\,$m\,s^{-2}$ in the LASCO/C2 field-of-view. \emph{Bottom right:} The automated CACTus velocities determined along the angular span of the CME, with a corresponding box-and-whisker plot to highlight the median (548\,$km\,s^{-1}$) and interquartile range.}
\label{20000102_kins_plots}
\end{figure}

In coronagraph images CMEs are observed as outwardly moving clouds of plasma, that appear relatively bright against the background corona (see the example in Fig.\,\ref{20000102_four}). Different methods for thresholding the CME brightness in such images have been employed by different catalogs. This is in order to detect their appearance and track their motion through the field-of-view, leading to a determination of the CME kinematics and morphology. However, each method suffers from drawbacks and, as such, the resulting CME catalogs can vary significantly in their characterisations and measurements of each event.

\subsection{CORIMP Automated Catalog}
\label{sect_corimp}

The CORIMP\footnote{\href{http://alshamess.ifa.hawaii.edu/CORIMP/}{http://alshamess.ifa.hawaii.edu/CORIMP}} catalog was developed with a method of dynamic signal separation and multiscale edge detection to overcome certain drawbacks of previous detection and tracking methods that rely on running-difference images. The CME signal is separated from the more quiescent streamers and coronal structures in LASCO/C2 and C3 images, such that a multiscale filtering technique may be used to suppress noise in order to characterise the CME structure and track its motion. A spread of heights of the CME front is then measured across its angular span. A type of cleaning algorithm is applied to the height-time measurements before the kinematics are determined. This is required to overcome cases where pixels along the CME front edges at a specific position angle are not correctly identified, but rather a pixel corresponding to core material behind the CME front is measured, which causes unnecessary scatter in the height-time datapoints. The cleaning algorithm works by stepping along the height-time measurements at each position angle and requiring that only sequentially increasing heights are plotted. This helps remove the detections of trailing CME material from the height-time plots such that only the kinematics of the main CME front are determined. The kinematics are then derived in three ways: using a Savitzky-Golay filter \citep{Savitzky-Golay1964}, a quadratic fit (second-order polynomial), and a linear fit (straight-line). The importance of a robust method for determining the kinematics of a transient event is discussed in \cite{2013A&A...557A..96B}, wherein the often-used method of 3-point Lagrangian interpolation and associated error-propagation were shown to behave counter-intuitively and provide misleading kinematic results. It was shown that the Savtizky-Golay smoothing filter performs well on CME height-time data, through its use of a kernel-based estimation of a local polynomial regression, that also directly computes the derivatives of the parameters to provide the kinematics. The number of data points either side of the case point to be included in the filter is also specified, being optimised for the cadence of the data. The three automated fitting techniques in CORIMP are useful for comparing with the other catalogs, and offer a degree of freedom for the user to chose the output most appropriate to their interests (e.g., a linear fit can provide an average CME speed over the observed interval, while the Savitzky-Golay filter can reveal changes to the acceleration in that interval).

\subsection{CACTus Automated Catalog}

The CACTus\footnote{\href{http://sidc.oma.be/cactus/}{http://sidc.oma.be/cactus/}} catalog was the first automated CME detection algorithm, in operation since 2004. It is based upon the detection of CMEs as bright ridges in time-height slices at each angle around a coronagraph image. A running-difference technique is applied to each image, before it is transformed into Sun-centred polar coordinates. These images are then re-binned and the separate C2 and C3 fields-of-view are combined. These are then stacked in time, and for each angle the corresponding time-height slice undergoes a modified Hough transform for detecting intensity ridges across it. Thresholding the most significant ridges filters out the progression of CMEs, with the variables for each ridge characterised by onset time, speed, and position angle. The median speed across the angular span of each event is quoted as the CME speed.

The running-difference cadence, the ridge intensity threshold, and the imposed limit on how many frames a CME may exist, all affect how successful the automated detection can be. However, \citet{2004A&A...425.1097R} show the algorithm to be robust in reproducing the detections of a human user by directly comparing with the CDAW catalog. The main drawback of the CACTus catalog is that it cannot resolve CME acceleration, since the Hough transform thresholds the ridges as straight lines whose slopes provide a constant speed. The speed itself may also be an underestimate since it is a median across the span of the CME, although having a spread of speed measurements across position angles is an advantage of CACTus (see the example speed plot at the bottom right of Fig.\,\ref{20000102_kins_plots}). However, it is sometimes possible that the angular spans are over-estimated since side outflows in the images are enhanced by the running-difference and may include streamer deflections. It is also difficult to distinguish when one CME has fully progressed from the field-of-view and another CME has entered it, so in some cases trailing portions of a CME are detected as separate events.

\subsection{SEEDS Automated Catalog}

The SEEDS\footnote{\href{http://spaceweather.gmu.edu/seeds/}{http://spaceweather.gmu.edu/seeds/}} catalog employs an automated CME detection algorithm for tracking an intensity thresholded CME front in running-difference images from LASCO/C2. The images are unwrapped into Sun-centred polar coordinates, and a normalised running-difference technique is applied (such that the mean intensity of the new image is effectively zero). The pixel intensities (positive values only) are then summed along angles and thresholded at a certain number of standard deviations above the mean intensity. This determines the ``core angles" of the CME, and a region growing technique based on a secondary threshold of intensities in the rest of the image is applied to open the angular span to include the full CME. An issue arises when streamer deflections occur that offset the region growing technique and overestimate the CME angular width. An intensity average across the angles within the span of the CME is then determined, and where the forward portion of this intensity profile equals half its maximum value is taken as the CME height. The speed and acceleration are determined from the heights through consecutive images and these results are output with the CME position angle and angular width in the SEEDS catalog (see the example height-time plot at the top right of Fig.\,\ref{20000102_kins_plots}).

Along with the issues of streamer deflections and the tracking being limited to only the C2 field-of-view, the choice of the ``Half-Max-Lead" as the CME height is dependant on the overall CME brightness, and thus any brightness change during its propagation will affect this measurement. This adds to the error on the height-time profile, which affects the accuracy of the derived speed and acceleration.

\subsection{CDAW Manual Catalog}

The CME catalog hosted at the CDAW Data Center\footnote{\href{http://cdaw.gsfc.nasa.gov/CME_list}{http://cdaw.gsfc.nasa.gov/CME\_list}}  grew out of a necessity to record a simple but effective description and analysis of each event observed with LASCO \citep{2009EM&P..104..295G}. The catalog is wholly manual in its operation, with a user tracking the CME through C2 and C3 running-difference images and producing a ``point-\&-click" height-time plot of each event. A linear fit to the height-time profiles provides a 1st-order estimate for the plane-of-sky speed, and a quadratic fit provides a 2nd-order speed fit and an acceleration for the event. The central position angle and angular width of the CME are also deduced from the images, and the event is flagged as a halo if it spans 360$^{\circ}$, partial halo if it spans $\ge$\,120$^{\circ}$, and wide if it spans $\ge$\,60$^{\circ}$. The catalog itself lists each CME's first appearance in C2, central position angle, angular width, linear speed, 2nd-order speed at final height, 2nd-order speed at 20~R$_{\odot}$, acceleration, mass, kinetic energy, and measurement position angle (the angle along which the heights of the CME are determined; see the example height-time and speed-height plots in the left of Fig.\,\ref{20000102_kins_plots}). While the human eye is supremely effective at distinguishing CMEs in coronagraph images, errors may be introduced to the manual cataloging procedure through the biases of different operators; for example, in deciding how the images are scaled, where along the CME the heights are measured, or whether a CME is worth including in, or discarding from, the catalog.

\section{CME Event Sample and Catalog Results}
\label{sect_events}

A selection of CMEs from the SOHO/LASCO data was chosen in the analysis of \cite{2009A&A...495..325B}, wherein multiscale methods of edge detection and a resulting ellipse characterisation of the CME front were used to track its apex. These events were chosen based on their varying styles of eruption and appearance, in order to compare with the measurements of the manual CDAW catalog (see images of each CME in Fig.\,5 of \citealt{2009A&A...495..325B}). They exhibit various forms that typical CMEs in coronagraph observations can take, to serve as examples of how the detection and characterisation algorithms fare on each, and how well their varying kinematic trends are revealed. The images were not differenced so that the problematics effects of spatiotemporal cross-talk were avoided. The uncertainties on the height measurements were quantified by the multiscale filter size and subsequent ellipse-fitting, and propagated into the kinematics via numerical differentiation using 3-point Lagrangian interpolation. However, it has since been demonstrated by \cite{2013A&A...557A..96B} that this method for deriving kinematics is not wholly reliable, and other approaches must be considered as discussed in Section\,\ref{sect_corimp} above. Following the development of the automated CORIMP algorithms, these events are now revisited in this new catalog and directly compared with the other automated CACTus and SEEDS catalogs, the manual CDAW catalog, and the results of \cite{2009A&A...495..325B}. In each case below, the tabled information and kinematic plots are reproduced directly from the online catalogs, and not rescaled or otherwise manipulated, for a fair comparison.


\begin{table}[!t]
\begin{tabular}{l*{5}{c}r}
\multicolumn{6}{c}{} \\
CME Date \& Start Time & Catalog              & CPA [$deg.$] & Width [$deg.$] & Lin. Speed [$km\,s^{-1}$] & Accel. [$m\,s^{-2}$]  \\
\hline
\hline
2000 Jan. 02 $\sim$06:06\,UT & CORIMP         & 250   & 81$^{83}$   & 454$^{743}$ &  1$_{-17}^{14}$  \\
(Arcade eruption) & CACTus          & 250 & 106 & 548$_{231}^{744}$ &     \\
& SEEDS        & 257 & 96 & 292 & 18.6  \\
& CDAW     & 253 & 107 & 603 & 21.6 \\
\hline
2000 Apr. 18 $\sim$14:54\,UT & CORIMP      & 210 & 98 & 431$^{537}$ & 4$_{-11}^{15}$   \\
(Gradual CME) & CACTus     & 198 & 102 & 463$_{227}^{744}$ &     \\
& SEEDS        & 195 & 108 & 338 & 17.7    \\
& CDAW         & 195 & 105 & 668 & 23.1    \\
\hline
2000 Apr. 23 $\sim$12:54\,UT & CORIMP   & 287 & 119$^{125}$ & 836$^{1706}$ & -11$_{-154}^{50}$       \\
(Impulsive CME) & CACTus  & 144 & 360 & 1114$_{245}^{1849}$ &            \\
& SEEDS  & 275 & 130 & 594 & -8.5        \\
& CDAW      & 281 & 360 & 1187 & -48.5 \\
\hline
2001 Apr. 23 $\sim$12:39\,UT & CORIMP   & 232 & 72$^{74}$ & 187$^{283}$ & 3$_{-13}^{15}$ \\
(Faint CME) & CACTus  & 231 & 88 & 459$_{315}^{602}$ &  \\
& SEEDS    & 224 & 77 & 408 & -46.6  \\
& CDAW      & 228 & 91 & 530 & -0.7 \\
\hline
2002 Apr. 21 $\sim$01:26\,UT & CORIMP   & 235 & 154$^{177}$ & 1129$^{2300}$ & 61$_{-619}^{345}$ \\
(Fast CME) & CACTus   & 322 & 352 & 1103$_{298}^{1913}$ &  \\
& SEEDS    & 250 & 186 & 703 & 31.8 \\
& CDAW      & 282 & 360 & 2393 & -1.4 \\
\hline
2004 Apr. 1 $\sim$23:04\,UT & CORIMP   & 58 & 42$^{44}$ & 401$^{502}$ & 2$_{-22}^{18}$ \\
(Slow CME) & CACTus  & 60 & 70 & 485$_{244}^{829}$ & \\
& SEEDS     & 60 & 59 & 261 & 19.7 \\
& CDAW      & 59 & 79 & 460 & 7.1 \\

\end{tabular}
\caption{Catalog measurements of a sample of CMEs observed by SOHO/LASCO. CME Date \& Start Time refers to the first observation of the CME in LASCO. CPA refers to the central position angle of the CME. Width refers to the angular span, or opening angle, of the CME. Lin. Speed is the derived speed of the CME using a linear fit to the height-time measurements. Accel. is the derived acceleration of the CME using a second-order fit to the height-time measurements. Note, some values have a corresponding maximum and/or minimum ($x_{min}^{max}$) as specified in the respective catalogs.}
\label{event_table}
\end{table}


\subsection{Arcade eruption: 2000 January 2}
\label{sect_20000102}

\begin{figure}[t]
\centerline{\includegraphics[scale=0.5, trim=20 70 0 140, clip=true]{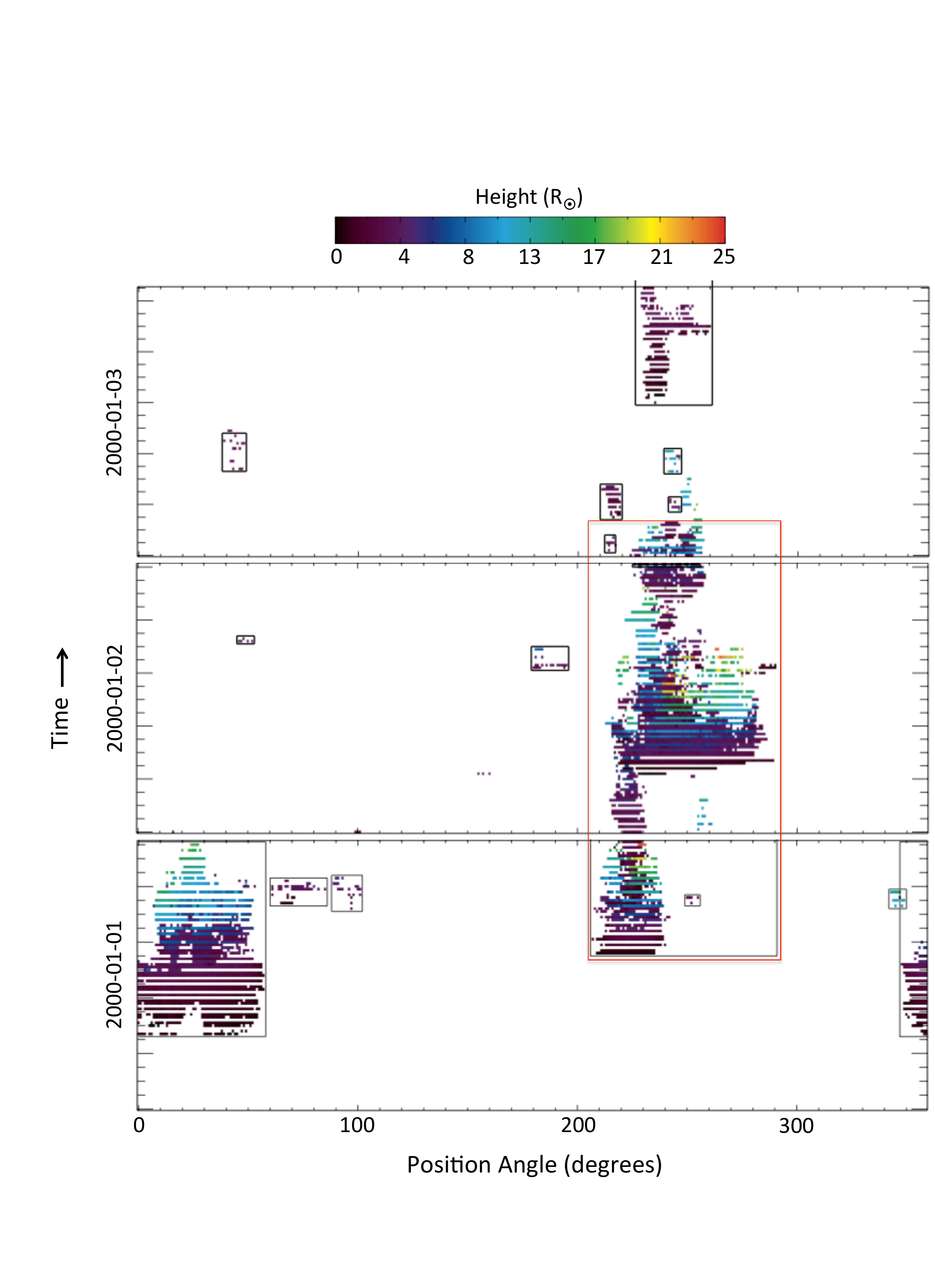}}
\caption{Daily detection stacks reproduced from the CORIMP CME catalog for the SOHO/LASCO observations in the date range 2000~Jan.~01\,--\,03. These stacks are generated from the automatically measured CME front heights at every position angle in an image, stacked in time. CMEs appear as groupings of colour-graded pixels, as indicated by the boxed regions. The overlapping CMEs in this time interval are boxed in red, corresponding to the height-time profiles in Fig.\,\ref{20000102_corimp_kinspd} (that have been put through the cleaning algorithm).}
\label{pa_total}
\end{figure}

\begin{figure}[t]
\centerline{\includegraphics[width=\linewidth]{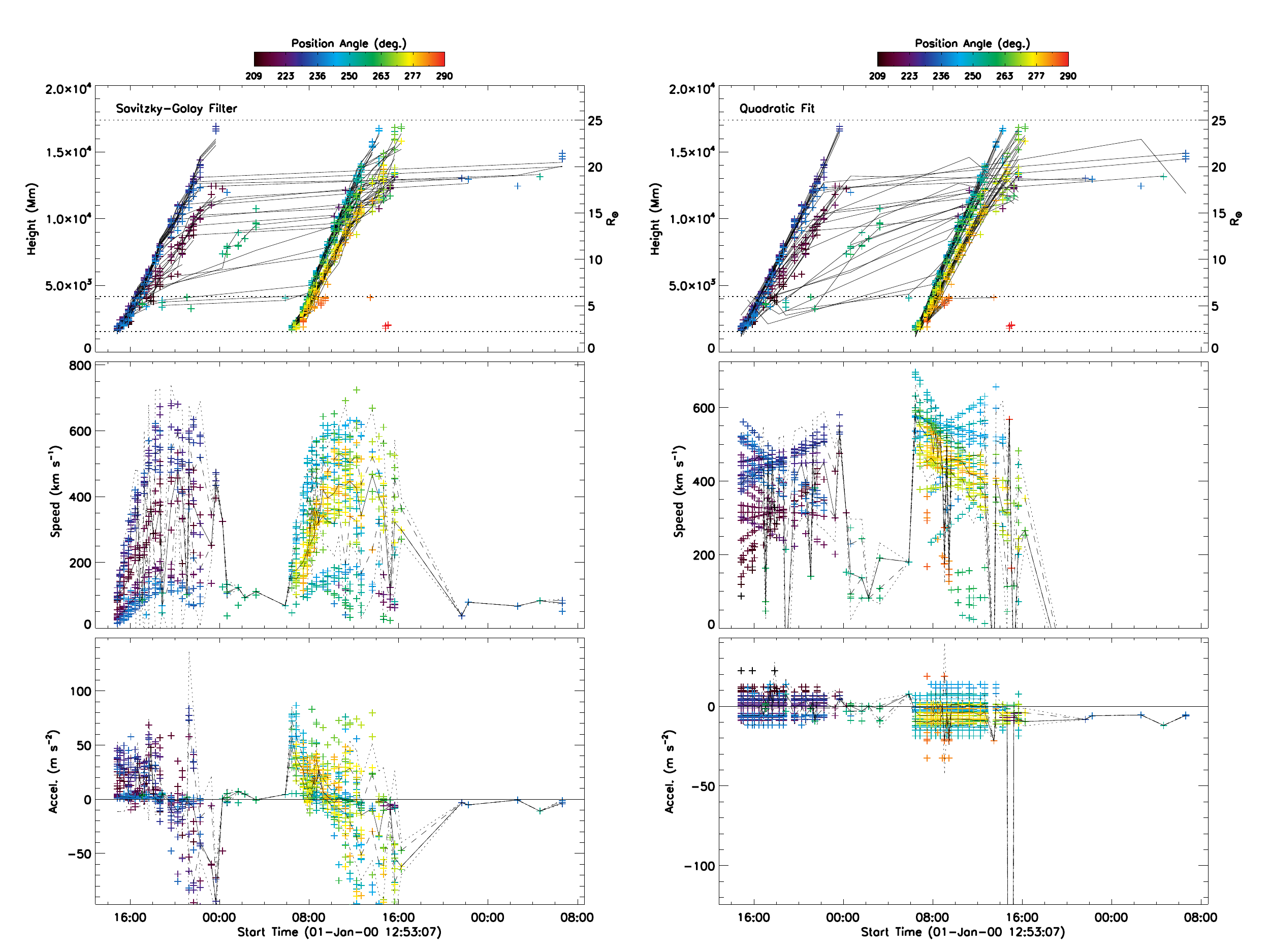}}
\caption{Kinematic plots of the 2000\,Jan.\,02 CME from the automatic detection and tracking in the CORIMP catalog. The top plots show the height-time measurements with a colorbar to indicate the angular span of the data points, and solid black lines to indicate the fitting. The middle and bottom plots show the speed and acceleration profiles of the CME with the median (solid line), interquartile range (inner dashed lines) and upper and lower fences (outer dashed lines) over-plotted. The left plots are determined by a Savitzky-Golar filter applied to the height-time measurements with a 7-point moving window, while the right plots are determined with a second-order quadratic fit.}
\label{20000102_corimp_kinspd}
\end{figure}

The CME that erupted off the southeast limb of the Sun on 2000~Jan.~02 from $\sim$06:06\,UT in LASCO exhibited an arcade-type structure consisting of multiple bright loops. CORIMP identified the bulk of the CME through the LASCO field-of-view to $\sim$24\,$R_\odot$. However, this CME may be deemed the third in a series of four CMEs that occurred in succession off the southeast limb, that CORIMP failed to separate due to their spatial and temporal overlap (essentially a smaller CME in between two large ones connects their detections along with a fourth smaller one afterwards, as seen in the CORIMP CME detection stacks reproduced in Fig.\,\ref{pa_total} - see \citealt{2012ApJ...752..145B} for details). This therefore serves as an example of the need to inspect the catalog output before trusting the quoted values listed in Table\,\ref{event_table}. The CORIMP height-time measurements (in the time range $\sim$06:00\,--\,16:00\,UT in Fig.\,\ref{20000102_corimp_kinspd}) reveal a non-linear trend indicative of an early acceleration that the Savitzky-Golay filter determines decreases from a maximum of $\sim$50\,$m\,s^{-2}$ to 0\,$m\,s^{-2}$, as the maximum speed levels off in the range of $\sim$500\,--\,600\,$km\,s^{-1}$. This is consistent with the measurements of \cite{2009A&A...495..325B} shown in their Fig.\,6. The quadratic (and linear) fits in CORIMP agree with a maximum speed in this range of $\sim$500\,--\,600\,$km\,s^{-1}$ and an acceleration in the range of approximately $\pm$20\,$m\,s^{-2}$. CACTus determined a linear speed of 548\,$km\,s^{-1}$ (in the range 231\,--\,744\,$km\,s^{-1}$). SEEDS determined a linear speed of 292\,$km\,s^{-1}$ and an acceleration of 18.6\,$m\,s^{-2}$ (in the C2 field-of-view). And CDAW determined a linear speed of 603\,$km\,s^{-1}$ and an overall acceleration of 21.6\,$m\,s^{-2}$. These catalog measurements are listed in Table\,\ref{event_table}. (Note that the slightly lower angular width in CORIMP is due to the exclusion of part of the questionable streamer deflection/interaction along the southern flank of the CME.) Therefore, by inspection, the results of the CORIMP CME catalog are in agreement with the other catalogs and manual analysis of this event, and CORIMP is deemed robust albeit unreliable at separating overlapping events.

\subsection{Gradual CME: 2000 April 18}

\begin{figure}[t]
\centerline{\includegraphics[width=\linewidth]{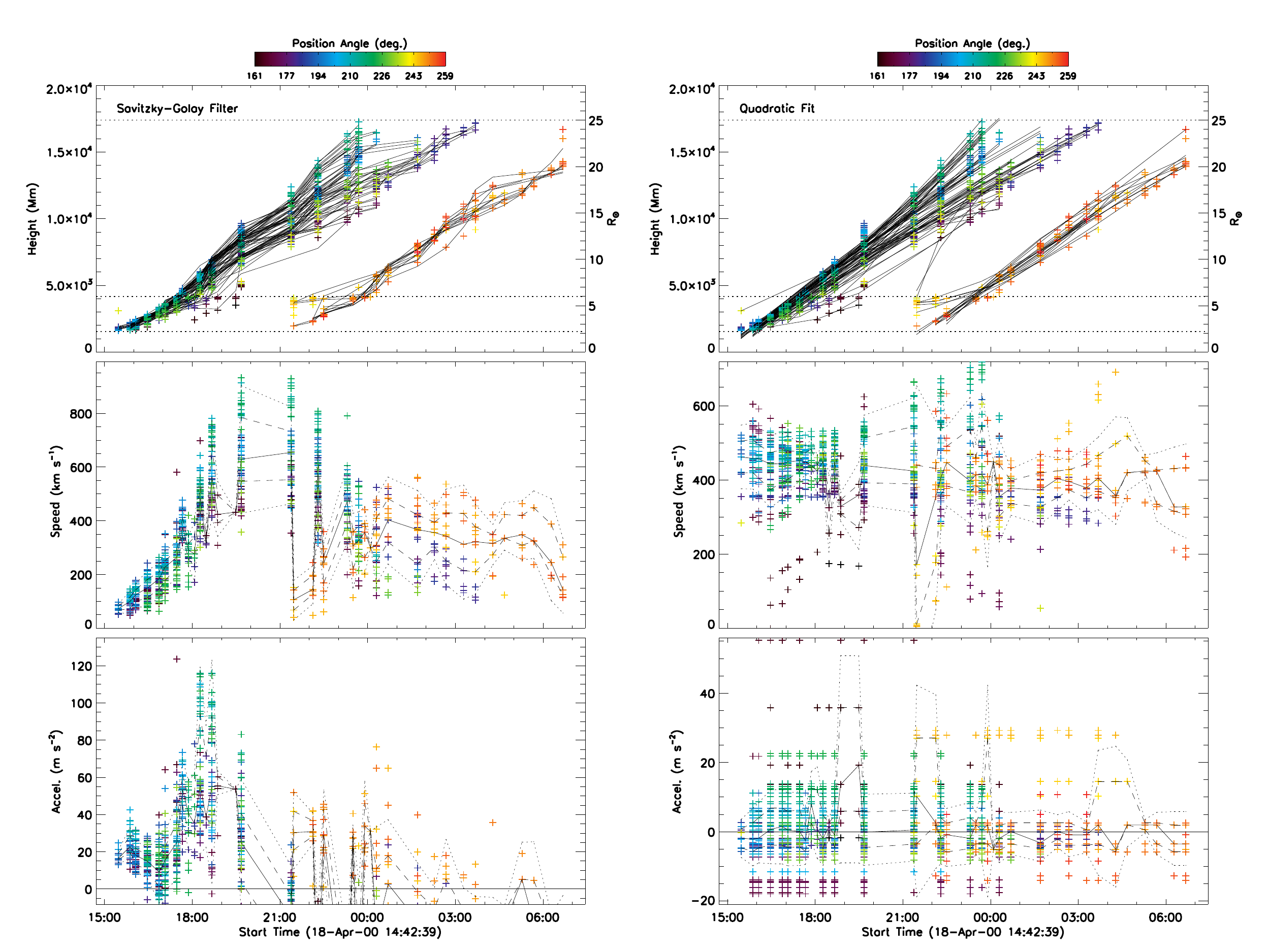}}
\caption{Kinematic plots of the 2000\,Apr.\,18 CME from the automatic detection and tracking in the CORIMP catalog, as in Fig.\,\ref{20000102_corimp_kinspd}.}
\label{20000418_corimp_kinspd}
\end{figure}

The CME that erupted off the south limb of the Sun on 2000~Apr.~18 from $\sim$14:54\,UT in LASCO exhibited a typical 3-part structure of leading CME front, cavity and bright core. CORIMP identified the bulk of the CME through the LASCO field-of-view to $\sim$25\,$R_\odot$, though it did not detect a southern portion of the faint CME front in the latter C3 observations. A western portion of material also erupted as a delayed part of the northern flank of the CME, that appears as a somewhat secondary height-time profile in the CORIMP kinematic plots in Fig.\,\ref{20000418_corimp_kinspd} (at position angles $\sim$250$^{\circ}$ in the redder end of the colorbar). The CORIMP height-time measurements reveal a non-linear trend indicative of an early acceleration that the Savitzky-Golay filter determines to be approximately 20\,$m\,s^{-2}$ as the speed increases to over 400\,$km\,s^{-1}$ before the data gap in the LASCO/C3 images between 19:42 and 21:24\,UT causes a large scatter in the derived kinematics (e.g., an artificial acceleration peak of $>$100\,$m\,s^{-2}$). The initial increasing speed profile up to a maximum in the range $\sim$600\,--\,800\,$km\,s^{-1}$ by $\sim$20:00\,UT agrees with that of \cite{2009A&A...495..325B} as shown in their Fig.\,7. The quadratic (and linear) fits in CORIMP are not as prone to the scattering effects of the data gap, and thus derive a slightly lower maximum speed range of $\sim$500\,--\,550\,$km\,s^{-1}$ and an acceleration in the range of approximately $\pm$15\,$m\,s^{-2}$. CACTus determined a linear speed of 463$\,km\,s^{-1}$ (in the range 227\,--\,744\,$km\,s^{-1}$). SEEDS determined a linear speed of 338$\,km\,s^{-1}$ and an acceleration of 17.7\,$m\,s^{-2}$ (in the C2 field-of-view). And CDAW determined a linear speed of 668$\,km\,s^{-1}$ and an overall acceleration of 23.1\,$m\,s^{-2}$. Therefore, by inspection, all sets of results are in agreement for this event.

\subsection{Impulsive CME: 2000 April 23}

\begin{figure}[t]
\centerline{\includegraphics[width=\linewidth]{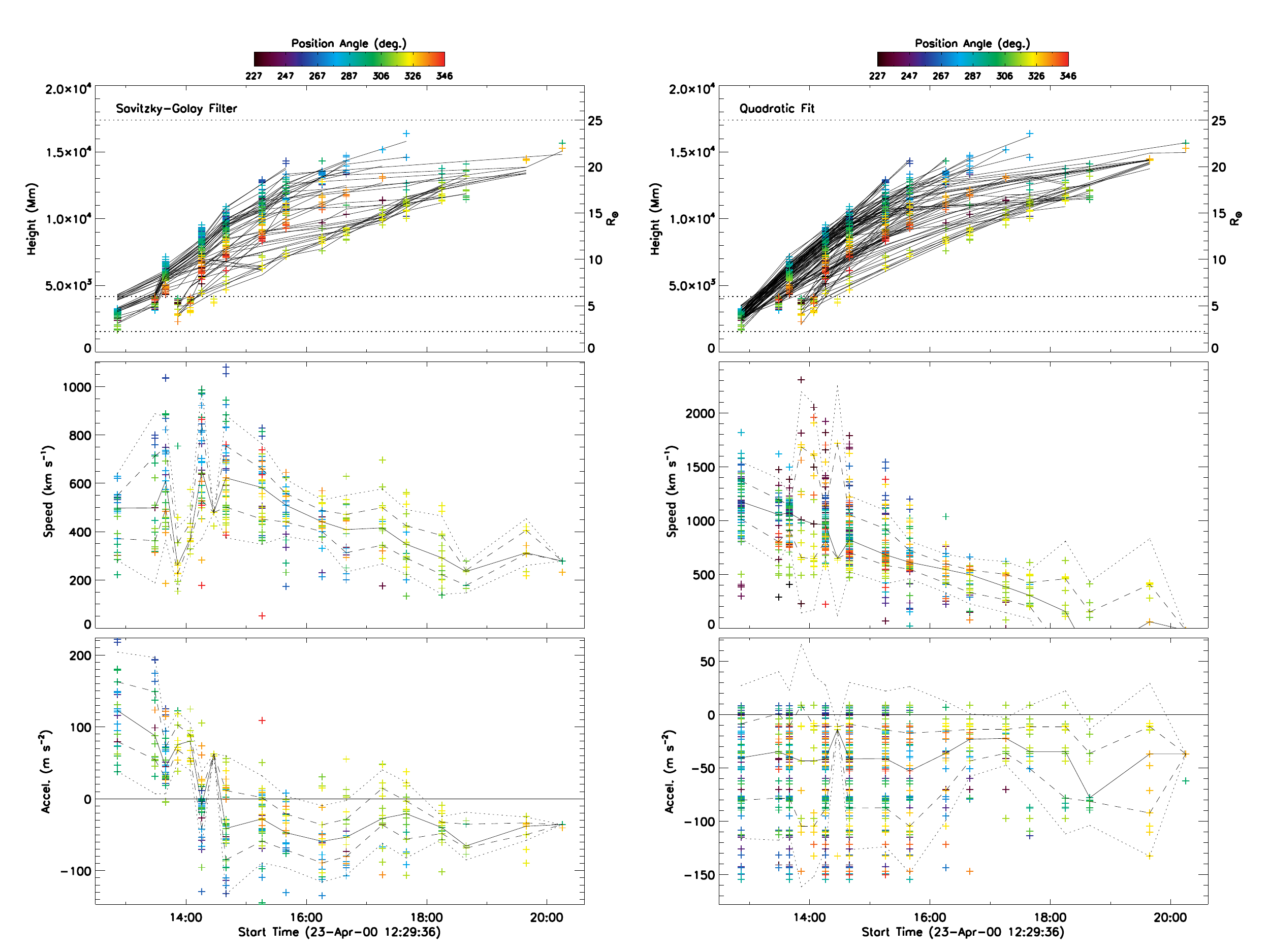}}
\caption{Kinematic plots of the 2000\,Apr.\,23 CME from the automatic detection and tracking in the CORIMP catalog, as in Fig.\,\ref{20000102_corimp_kinspd}.}
\label{20000423_corimp_kinspd}
\end{figure}

The large and fast CME that erupted off the west limb of the Sun on 2000~Apr.~23 from $\sim$12:54\,UT in LASCO underwent a hugely impulsive acceleration as it exploded into the corona. CORIMP identified the bulk of the CME through the LASCO field-of-view to $\sim$20\,$R_\odot$ after which the CME front became too faint. Strong streamer deflections occurred to the north and south flanks of the CME, with very faint material visible as a full halo or shock around the east limb separate to the bulk flux-rope structure in the west. The CORIMP height-time measurements (Fig.\,\ref{20000423_corimp_kinspd}) reveal an initial acceleration that the Savitzky-Golay filter determines to be $\gtrsim$150\,$m\,s^{-2}$ dropping quickly to a range of approximately $-100$ to 0\,$m\,s^{-2}$, as the speed decreases from $\sim$1000 to 500\,$km\,s^{-1}$; though this is an underestimate since the filter overly smoothes the relatively under-sampled height-time measurements. The quadratic fits in CORIMP better handle this data and derive an initial speed range of $\sim$1200\,--\,1500\,$km\,s^{-1}$, while the linear fits derive an initial speed range of $\sim$1000\,--\,1200\,$km\,s^{-1}$, which are consistent with the measurements of \cite{2009A&A...495..325B} shown in their Fig.\,8. The resulting deceleration is determined to have a median of approximately $-50\,m\,s^{-2}$, reaching as low as $-150\,m\,s^{-2}$. CACTus determined a linear speed of 1114$\,km\,s^{-1}$ (in the range 245\,--\,1849\,$km\,s^{-1}$). SEEDS determined a linear speed of 594\,$km\,s^{-1}$ and a deceleration of $-8.5\,m\,s^{-2}$ (in the C2 field-of-view). And CDAW determined a linear speed of 1187$\,km\,s^{-1}$ and an overall deceleration of $-48.5\,m\,s^{-2}$. Therefore, by inspection and careful consideration of the low sampling of the event, the results of the CORIMP CME catalog are in agreement with the corresponding results of the other catalogs and manual analysis.

\subsection{Faint CME: 2001 April 23}
\label{sect_20010423}

\begin{figure}[t]
\centerline{\includegraphics[width=\linewidth]{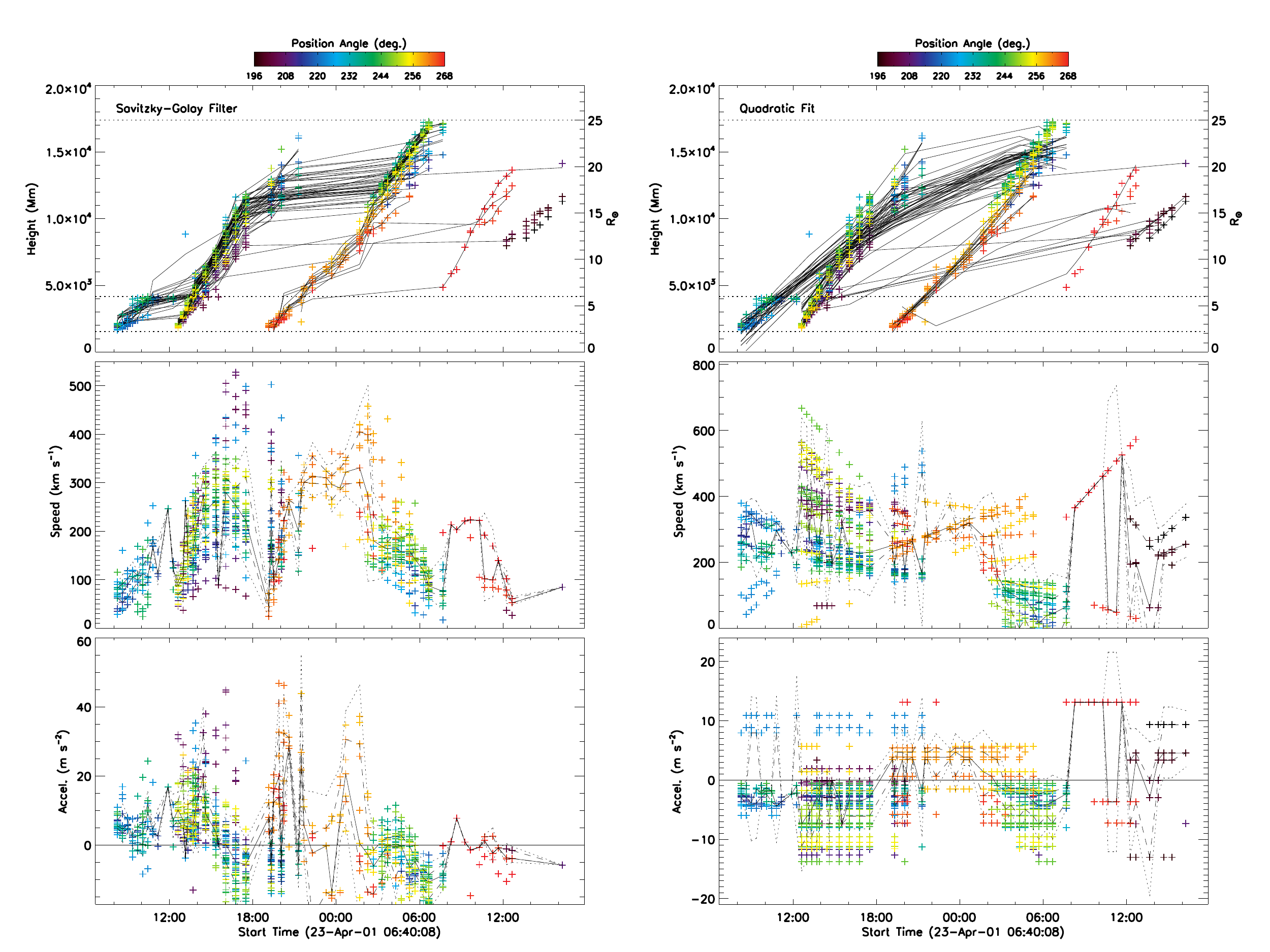}}
\caption{Kinematic plots of the 2001\,Apr.\,23 CME from the automatic detection and tracking in the CORIMP catalog, as in Fig.\,\ref{20000102_corimp_kinspd}.}
\label{20010423_corimp_kinspd}
\end{figure}

The CME that erupted off the southwest limb of the Sun on 2001~Apr.~23 from $\sim$12:54\,UT in LASCO appeared relatively faint behind multiple streamers in the line-of-sight, some of which deflected especially along the southern flank of the CME. CORIMP identified the bulk of the CME through the LASCO field-of-view to $\sim$20\,$R_\odot$ after which the CME front became too faint. However, this CME was the first of two that occurred in close succession off the southwest limb, that CORIMP failed to separate due to their spatial and temporal overlap (plus some ejecta ahead of this CME was detected from $\sim$08:16\,UT). Therefore the kinematic profiles must be inspected before trusting the quoted catalog values listed in Table\,\ref{event_table}. Investigating the relevant portion of the plots in Fig.\,\ref{20010423_corimp_kinspd}, in the time interval $\sim$12:00\,--\,18:00\,UT, the CORIMP height-time measurements reveal an initial acceleration that the Savitzky-Golay filter determines to be $\sim$10\,$m\,s^{-2}$ dropping to scatter about zero as the maximum speed levels off at $\sim$350\,$km\,s^{-1}$; though this is an underestimate since the measurements are dominated by the material ahead of the CME front that the algorithm detected as part of the main event (from $\sim$08:00\,--\,12:00\,UT). The quadratic fits to the measurements are more dominated by the overall deceleration of the CME (approx. $-10\,m\,s^{-2}$ from 12:00\,UT onwards) as the speed drops from an initial range of $\sim$550\,--\,650\,$km\,s^{-1}$ (consistent with \citealt{2009A&A...495..325B} shown in their Fig.\,9) to $\sim$400\,$km\,s^{-1}$ by 18:00\,UT; though this appears biased to lower values by the overlapping measurements of the second CME. The linear fits are less trustworthy as they tend to fit across the two CMEs and preceding ejected material, resulting in the underestimated CORIMP linear speed in Table\,\ref{event_table}. CACTus determined a linear speed of 459$\,km\,s^{-1}$ (in the range 315\,--\,602\,$km\,s^{-1}$). SEEDS determined a linear speed of 408$\,km\,s^{-1}$ and a deceleration of $-46.6\,m\,s^{-2}$ (in the C2 field-of-view), however it failed to detect the CME front in the final frames which accounts for this erroneously large deceleration. And CDAW determined a linear speed of 530$\,km\,s^{-1}$ and an overall deceleration of $-0.7\,m\,s^{-2}$. Therefore, while all sets of results are found to be in agreement, there is again the issue of separating overlapping event kinematics.

\subsection{Fast CME: 2002 April 21}
\label{sect_20020421}

\begin{figure}[t]
\centerline{\includegraphics[width=\linewidth]{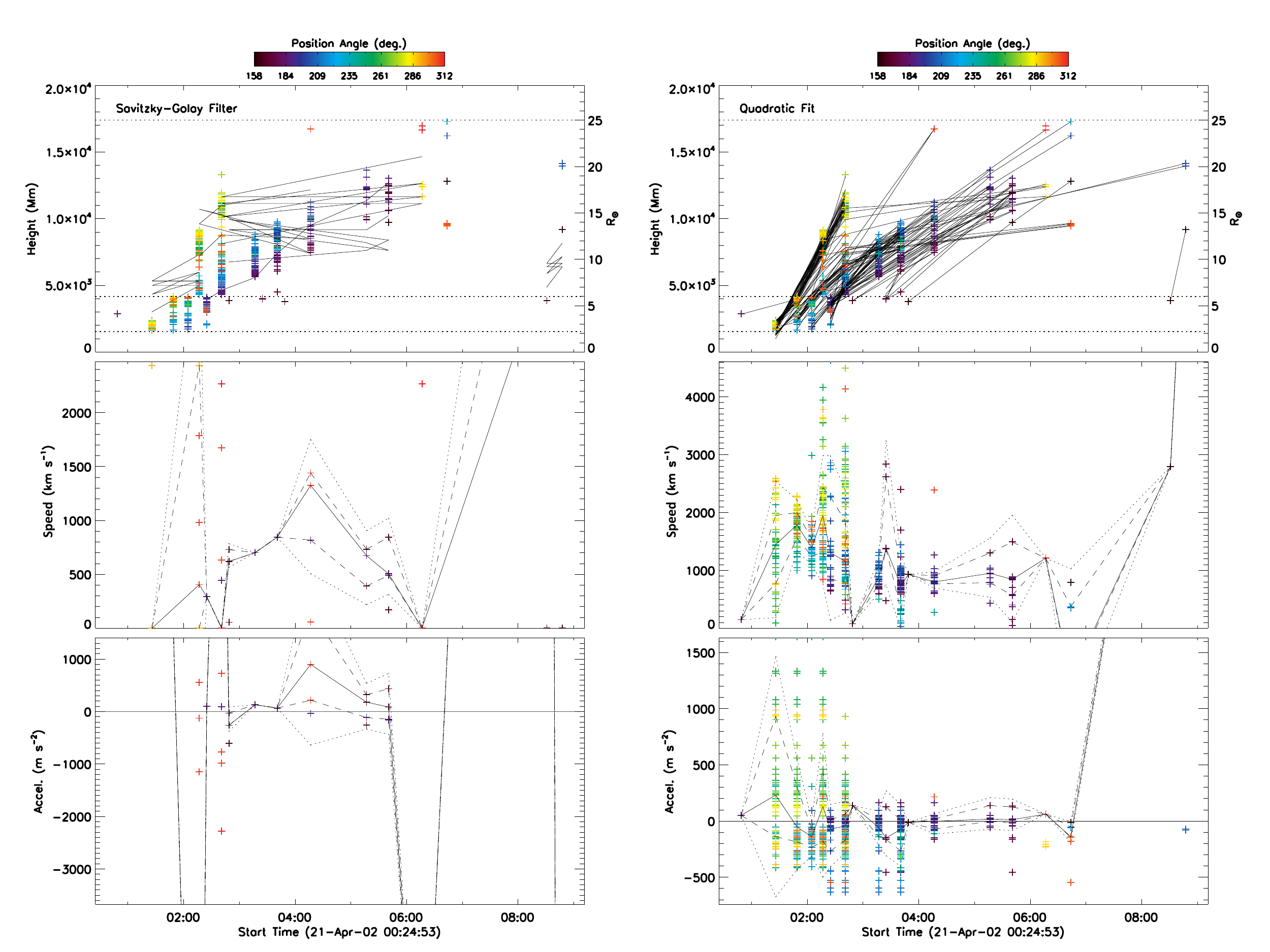}}
\caption{Kinematic plots of the 2002\,Apr.\,21 CME from the automatic detection and tracking in the CORIMP catalog, as in Fig.\,\ref{20000102_corimp_kinspd}.}
\label{20020421_corimp_kinspd}
\end{figure}

The CME that erupted off the west limb of the Sun on 2002~Apr.~21 from $\sim$01:27\,UT in LASCO propagated very fast through the field-of-view. CORIMP identified the bulk of the CME through the LASCO field-of-view to $\sim$17\,$R_\odot$ after which the CME front became too faint, and only the southern flank material continued to be detected. Figure\,\ref{20020421_corimp_kinspd} shows the CORIMP height-time measurements, which the Savitzky-Golay filter struggles to fit appropriately due to the small window-size available at each position angle (as the filter requires a minimum of 7 data points). The quadratic fits to the data reveal a high initial acceleration of $\gtrsim$1000\,$m\,s^{-2}$ followed by a deceleration in the range of approximately $-500$ to $0\,m\,s^{-2}$. The speed shows an initial range of $\sim$2000\,--\,2500\,$km\,s^{-1}$ possibly reaching $\sim$3000\,$km\,s^{-1}$ before dropping to $\sim$1000\,$km\,s^{-1}$, which is consistent with the measurements of \cite{2009A&A...495..325B} shown in their Fig.\,10. The linear fits also reveal an initial speed range of $\sim$2000\,--\,2500\,$km\,s^{-1}$ dropping to $\sim$1000\,$km\,s^{-1}$. CACTus determined a linear speed of 1103$\,km\,s^{-1}$ (in the range 298\,--\,1913\,$km\,s^{-1}$). SEEDS determined a linear speed of 703\,$km\,s^{-1}$ and an acceleration of $31.8\,m\,s^{-2}$ (in the C2 field-of-view), however these cannot be trusted as the CME front is only visible in two C2 frames. And CDAW determined a linear speed of 2393$\,km\,s^{-1}$ and an overall deceleration of $-1.4\,m\,s^{-2}$. Therefore, it is found that the Savitzky-Golay filter can be unreliable for characterising low-sampled events, but the quadratic and linear fits remain reliable and robust.

\subsection{Slow CME: 2004 April 1}

\begin{figure}[t]
\centerline{\includegraphics[width=\linewidth]{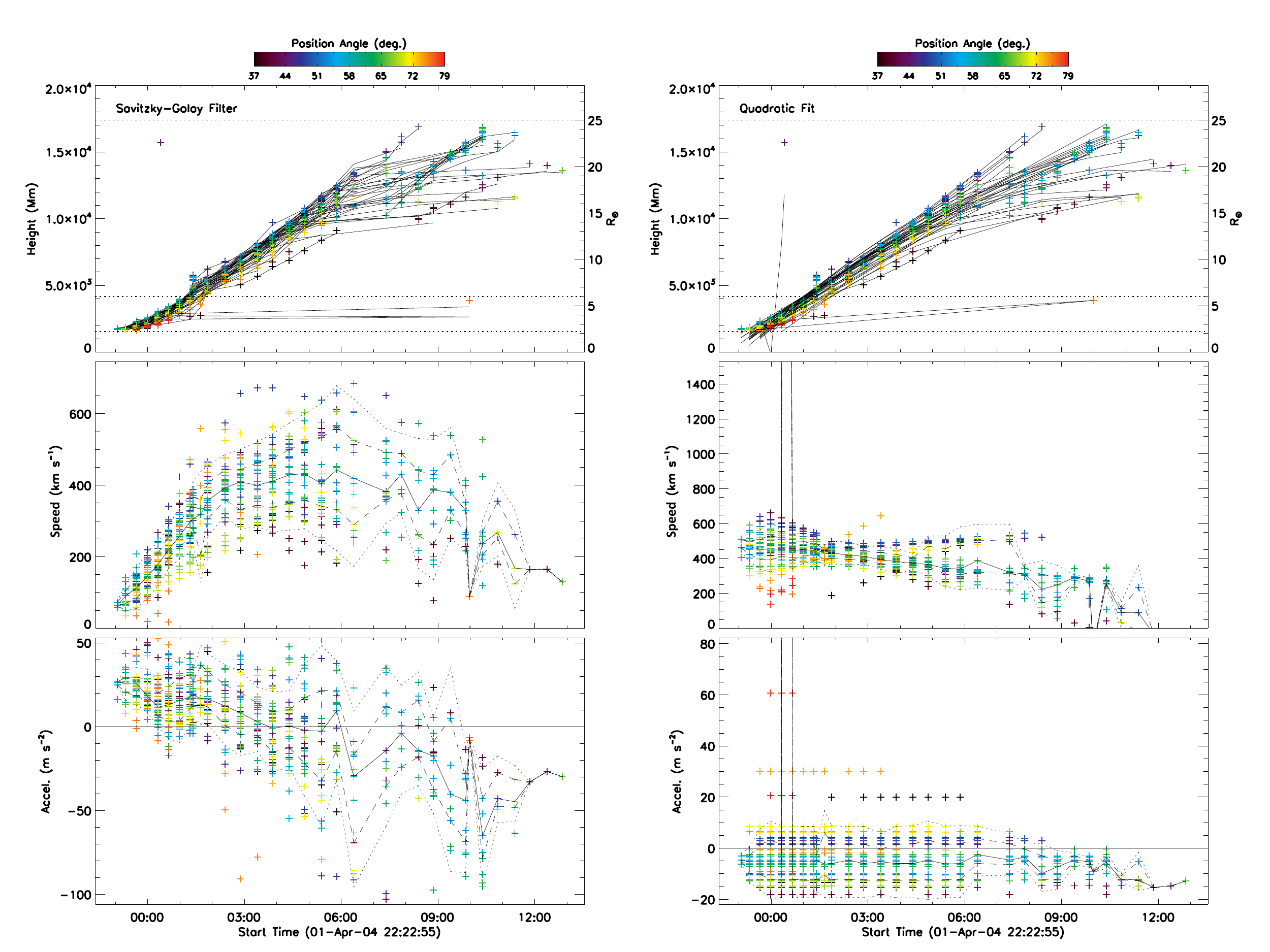}}
\caption{Kinematic plots of the 2004\,Apr.\,01 CME from the automatic detection and tracking in the CORIMP catalog, as in Fig.\,\ref{20000102_corimp_kinspd}.}
\label{20040401_corimp_kinspd}
\end{figure}

The CME that erupted off the northeast limb of the Sun on 2004~Apr.~01 from $\sim$23:05\,UT in LASCO, exhibited a clear flux-rope structure and propagated relatively slowly. CORIMP identified the bulk of the CME through the LASCO field-of-view to $\sim$20\,$R_\odot$ after which the CME front became too faint. Figure\,\ref{20040401_corimp_kinspd} shows the CORIMP height-time measurements, which are plentiful given the slow motion and clean detection of the event. These measurements reveal an initial acceleration that the Savitzky-Golay filter determines to be $\gtrsim$25\,$m\,s^{-2}$ dropping to 0\,$m\,s^{-2}$ by the time the CME reaches $\sim$15\,$R_\odot$ and the maximum speed levels off in the range $\sim$500\,--\,600\,$km\,s^{-1}$. The quadratic fits to the data reveal a bulk speed in the range $\sim$400\,--\,600\,$km\,s^{-1}$, with an overall deceleration of the CME of approximately $-5\,m\,s^{-2}$. The linear fits also produce a speed in this range. These results are consistent with the measurements of \cite{2009A&A...495..325B} shown in their Fig.\,11, though without reproducing the ``staggered" speed profile. CACTus determined a linear speed of 485$\,km\,s^{-1}$ (in the range 244\,--\,829\,$km\,s^{-1}$). SEEDS determined a linear speed of 261\,$km\,s^{-1}$ and overall acceleration of $19.7\,m\,s^{-2}$ (in the C2 field-of-view). And CDAW determined a linear speed of 460$\,km\,s^{-1}$ and an overall acceleration of $7.1\,m\,s^{-2}$. Therefore, by inspection, CORIMP and the other CME characterisations are in agreement for this event.

\section{Separating Multiple CME Detections via $K$-means Clustering}
\label{sect_clusters}

The case studies in the previous section highlight a key issue in the automatic detection, tracking and cataloging of CMEs: namely the difficulties in distinguishing between multiple events that occur close together in space and time. The events of Section\,\ref{sect_20000102} and \ref{sect_20010423} demonstrate how the CORIMP catalog can fail to separately characterise CMEs that a human user would label as distinct events (although even this can be a non-trivial task since projection effects can make it hard to determine if two CMEs are truly merging in space or simply overlapping on the plane-of-sky). The reason for this, is that the thresholds in place to identify the beginning of a new CME detection cannot readily determine the end of a previous CME detection whose trailing material overlaps the subsequent CME material on the plane-of-sky. Other observational factors can help a human user distinguish the two, such as the differing speeds, densities, brightness and cohesiveness of the structure. However, these differences are too subtle to employ in an automated algorithm that must be able to characterise all manner of CMEs with a large variety of such properties; and so the overlapping CMEs are classified as a single event. While it is still possible to investigate their separate kinematic trends, as in Section\,\ref{sect_20000102} and \ref{sect_20010423}, this is not ideal for accurately counting CMEs nor for reliably producing independent CME detection alerts. It therefore remains a challenge to employ some form of machine intelligence in cataloging CMEs, which can use the CME detection parameters to reveal instances when multiple CMEs occur together.

\begin{figure}[t]
\centerline{\includegraphics[scale=0.578, trim=0 95 0 50, clip=true]{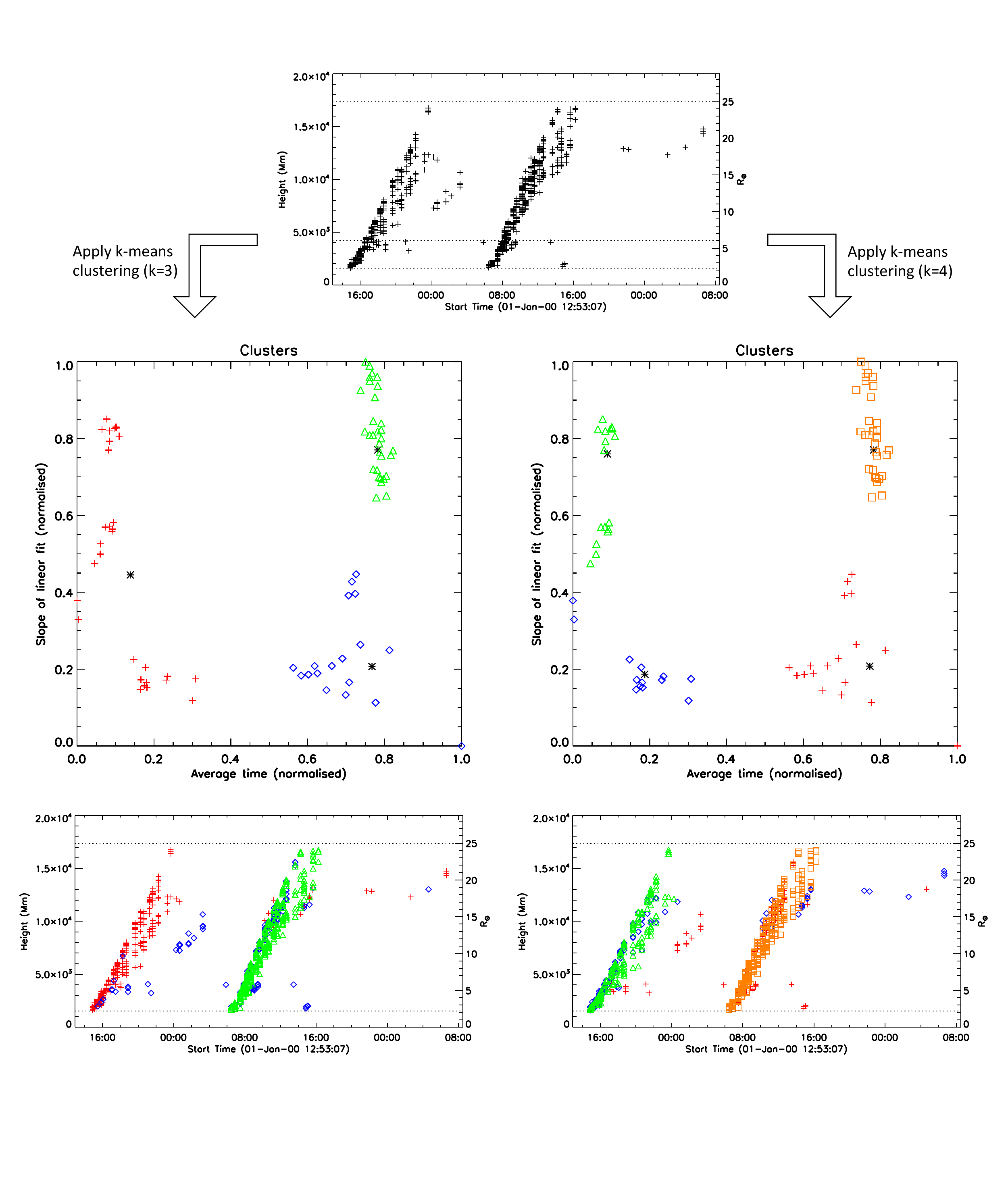}}
\caption{A $k$-means clustering algorithm applied to the height-time data on 2000\,Jan.\,01\,--\,03 (from Fig.\,\ref{20000102_corimp_kinspd}), in an effort to distinguish the multiple CME profiles that were detected as a single event due to their close proximity in space and time. \emph{Top plot:} The height-time measurements of the CME detections from the automated CORIMP catalog. (Note, this dataset has been put through a cleaning algorithm, discussed in Section\,\ref{sect_corimp}, that removes a lot of inner-core and trailing-material datapoints, thus making it easier to distinguish their separate profiles.) \emph{Middle plots:} The resulting clusters for the cases of $k$\,=\,3 (\emph{left}) and $k$\,=\,4 (\emph{right}), applied to the normalised parameters of the slope of a linear fit to, and the mean time of, the height-time profile at each position angle. The clusters are distinguished by different plot symbols and colours, with black asterisks to indicate the mean of each cluster. \emph{Bottom plots:} The resulting effort at separately distinguishing the CME height-time profiles.
}
\label{20000101_cluster_kins}
\end{figure}

\begin{figure}[t]
\centerline{\includegraphics[scale=0.578, trim=0 95 0 50, clip=true]{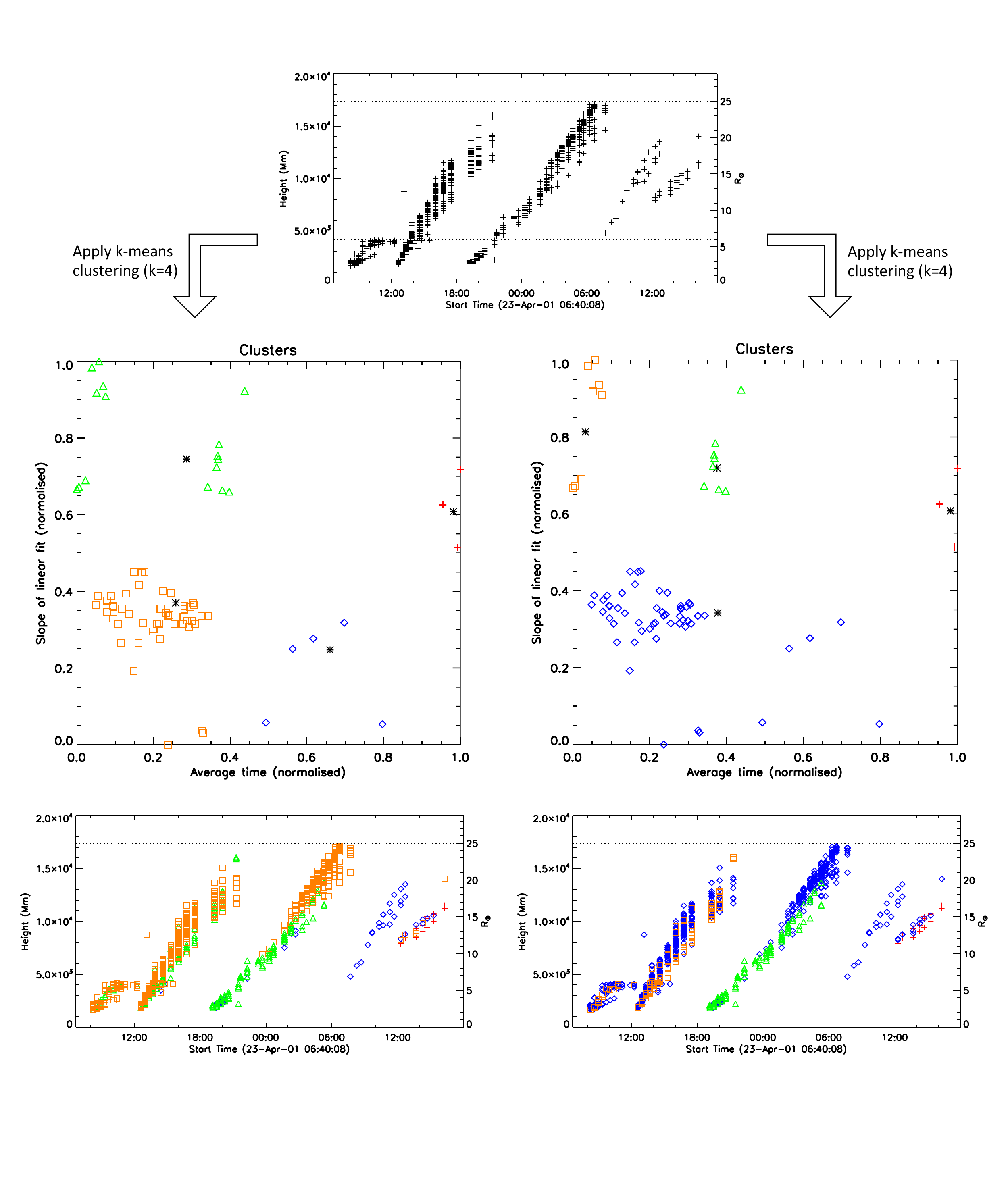}}
\caption{A $k$-means clustering algorithm applied to the height-time data on 2001\,Apr.\,23\,--\,24 (from Fig.\,\ref{20010423_corimp_kinspd}), in an effort to distinguish the multiple CME profiles, as in Fig.~\ref{20000101_cluster_kins} but for two cases of $k$\,=\,4.
}
\label{20010423_cluster_kins}
\end{figure}

An initial effort to do this has been made using a clustering algorithm in the field of unsupervised machine learning. Specifically, the method of $k$-means clustering was investigated, which works by partitioning $n$ observations into $k$ clusters that are distinguished by minimising the within-cluster sum of squares, i.e., using Euclidean distance as a metric on the parameter space. It is well suited to generating globular, non-hierarchical, non-overlapping clusters, and may be computationally fast if $k$ remains small. This approach could work with the parameters available in the CME detection analysis, such as the time, location/direction, size and speed of a CME. For example, the bulk of the overlapping CMEs may have different average propagation times as one may be proceeding slightly later in time than another. Or if the CMEs are propagating at different speeds, then a linear fit to each of their height-time profiles would have a different slope. Choosing these parameters of ``average time" and ``slope of a linear fit", determined at every position angle in the span of the event detection, it is possible to cluster the height-time measurements into separate CMEs.

The use of clustering techniques for separating overlapping CME height-time profiles is demonstrated in Figs.\,\ref{20000101_cluster_kins} and \ref{20010423_cluster_kins} for the two events discussed in Sections\,\ref{sect_20000102} and \ref{sect_20010423}, respectively. These figures show top plots of the $k$-means clustering algorithm (where $k$ is manually prescribed by the user) applied to the parameters of ``average time" and ``slope of a linear fit", where the means are plotted as black asterisks and the associated groups of points in separate clusters are plotted with different symbols and colours. The bottom plots of these figures show the corresponding height-time profiles that have been separated according to their clusters. In Fig.\,\ref{20000101_cluster_kins} the results are shown for both $k$\,=\,3 (left plots) and $k$\,=\,4 (right plots), to illustrate the effect of changing $k$. By inspection, the clustering algorithm works well at distinguishing the separate events. The bulk measurements of the case-study CME beginning at $\sim$06:06\,UT on 2000\,Jan.\,02 are quite well clustered, though some of its later C3 measurements are wrongly determined as part of a separate CME. For this event the $k$\,=\,3 case fares better at grouping the CMEs, while the $k$\,=\,4 case splits apart the profile of the first CME (shown as the green and blue datapoints in the bottom right plot of Fig.\,\ref{20000101_cluster_kins}). Similarly in Fig.\,\ref{20010423_cluster_kins} the clustering algorithms go some way towards distinguishing multiple CMEs in the event detection on 2001\,Apr.\,23, but there are datapoints that are wrongly classified, even for the two instances of $k$\,=\,4 shown for this event. These results highlight the difficulty in applying an automatic extraction of separate CME height-time profiles when detected so close together in space and time. Furthermore, there is an inherent limitation to $k$-means clustering by having to specify the number of clusters required from the data, which is not known a priori - especially not for an automated methodology such as in the CORIMP catalog. Further investigation into the parameters to be clustered, alternative clustering, or different machine learning algorithms, may produce better results.

\section{Conclusions}
\label{sect_conclusions}

As the wealth of coronagraph data and CME observations has increased dramatically since the launch of SOHO in 1995, it has become important to develop robust and reliable methods of detecting and tracking CMEs in white-light images. Since CMEs are faint and transient phenomena that prove difficult to consistently isolate from the background corona, manual inspection of the images is open to interpretation and prone to user-specific biases. Similarly, it is challenging to fix the criteria and thresholds necessary in a computerised methodology for automating this task, although advances have been made to achieve this and provide the benefit of having a self-consistent catalog of results. Efforts to both manually and automatically catalog CMEs have been discussed in Section\,\ref{sect_catalogs} with the aim of comparing how each fares in light of the newly developed CORIMP catalog, which was built to overcome some of the drawbacks of current catalogs. To this end, a selection of CMEs was chosen from a previous study by \cite{2009A&A...495..325B}, and the new results in the CORIMP catalog were investigated alongside the results of the automated CACTus and SEEDS catalogs and the manual CDAW catalog. 

In the previous study of \cite{2009A&A...495..325B}, the CMEs were characterised with the use of a multiscale edge-detection filter, whereby an ellipse was fitted to the isolated CME front and its apex tracked to produce height-time measurements. Since this approach avoided differencing the images, it was possible to quantify single-image uncertainties for the resulting height-time measurements, to be used for gauging a confidence interval on the derived CME kinematics. However, \cite{2013A&A...557A..96B} demonstrated that the often-used method of numerical differentiation using 3-point Lagrangian interpolation, and its associated error propagation, is not wholly reliable at deriving the true CME kinematics. This motivated the use of the Savitzky-Golay filter along with quadratic and linear fits to the height-time measurements in CORIMP, across the angular span of the CME such that the statistical spread in the kinematics of each event may better indicate the true underlying trends. It is therefore warranted to compare these new automatically-generated results with the outputs of the other catalogs.

The spread of measurements along the angular span of the CME proves more useful than choosing a single fixed apex of the CME, because it propagates as an impermanent, evolving structure that can undergo various rates of expansion across the plane-of-sky. The variety of events chosen here as a subset of the thousands in the LASCO data is enough to demonstrate this. Having the angular spread of kinematics also provides insight to the bulk motion of the CME as well as its flanks and front: with the angular extent indicating the flanks and the upper values on the velocities indicating the CME front (usually the fastest part of its structure). Therefore a greater amount of information is available on the overall CME motion.

The Savitzky-Golay filter provides an indication of the kinematic trends that a first or second-order fit cannot necessarily produce. Since this filter is applied in a moving-window on the datapoints, it can be problematic in cases of events with low-sampling (as in Section\,\ref{sect_20020421}), but otherwise performs very well at automatically quantifying the different phases of acceleration of a CME. Therefore the dynamics of the eruption may be better quantified and understood.  

While the robustness of the CORIMP catalog is clear (in so far as it can demonstrably produce results that are accurate and consistent across the data), there is a reliability issue that arises in cases of multiple CMEs that overlap in space and time. The problem with such cases is that another CME can erupt in the same direction as a previous one, close enough in time that the two detections are merged, as though the second CME were part of the trailing material of the first (as shown in Fig.\,\ref{pa_total}). The opposite problem to this is that harsher thresholds would split apart single CMEs into multiple events, especially large CMEs with substantial trailing material. Indeed such problems can affect all automated catalogs, such that CORIMP appears to suffer from the former issue, while CACTus and SEEDS suffer from the latter. A form of unsupervised machine learning was explored, by applying a $k$-means clustering algorithm to certain parameters in the CME detections, in an effort to distinguish overlapping events. While these first results shown in Section\,\ref{sect_clusters} are promising, they highlight the difficulty of the task, and warrant further investigation. For example, perhaps a form of supervised machine learning would fare better, if a substantial training set of correctly labelled event data was produced from the current database of results and used to train an intelligent algorithm. For now, this issue is only overcome by a manual inspection of the data, as highlighted in the events of Section\,\ref{sect_20000102} and \ref{sect_20010423}. In conclusion, any catalog should not be quoted blindly, as the thresholds cannot always distinguish the exact eruption that a user would isolate by eye. However, knowing this, CORIMP still offers the most rigorous details on the kinematics and morphologies of CMEs in a catalog to date, from which a user can infer a wealth of information.

\begin{acknowledgements}
     
The SOHO/LASCO data used here are produced by a consortium of the Naval Research Laboratory (USA), Max-Planck-Institut fuer Aeronomie (Germany), Laboratoire d'Astronomie (France), and the University of Birmingham (UK). SOHO is a project of international cooperation between ESA and NASA.
The CACTus CME catalog is generated and maintained by the SIDC at the Royal Observatory of Belgium.
The SEEDS CME catalog has been supported by NASA Living With a Star Program and NASA Applied Information Systems Research Program.
The CDAW Data Center CME catalog is generated and maintained by NASA and The Catholic University of America in cooperation with the Naval Research Laboratory.
The author acknowledges and thanks Huw Morgan (Aberystwyth University, Wales), Shadia Habbal (Institute for Astronomy, Hawaii), Peter Gallagher (Trinity College Dublin, Ireland) and Jackie Davies (RAL Space, UK) for their helpful and ongoing discussions regarding this work.
The author and editor thank two anonymous referees for their assistance in evaluating this paper.      
     
\end{acknowledgements}


\bibliographystyle{swsc}
\bibliography{references}  


\end{document}